\documentclass{article}
\usepackage[a4paper, portrait, margin=1in]{geometry}
\usepackage[T1]{fontenc}
\usepackage[utf8]{inputenc}
\usepackage{authblk}
\usepackage{subcaption}
\usepackage[belowskip=2pt,aboveskip=4pt, font={footnotesize}]{caption}

\usepackage{amssymb}
\usepackage{graphicx}
\usepackage{cite}
\usepackage[belowskip=2pt,aboveskip=4pt, font={footnotesize}]{caption}
\usepackage{fixltx2e}
\usepackage{multirow}
\usepackage{hhline}
\usepackage{array}
\newcolumntype{C}[1]{>{\centering\let\newline\\\arraybackslash\hspace{0pt}}m{#1}}

\begin{document}
\title{GHz-Band Integrated Magnetic Inductors}

\author[1]{Amal El-Ghazaly}
\author[2]{Robert M. White}
\author[1,2]{Shan X. Wang}

\affil[1]{Electrical Engineering, Stanford University, Stanford, CA 94305 USA}
\affil[2]{Materials Science and Engineering, Stanford University, Stanford, CA 94305 USA}

\maketitle

\begin{abstract}
The demand on mobile electronics to continue to shrink in size while increase in efficiency drives the demand on the internal passive components to do the same. Power amplifiers require inductors with small form factors, high quality factors, and high operating frequency in the single-digit GHz range. This work explores the use of magnetic materials to satisfy the needs of power amplifier inductor applications. This paper discusses the optimization choices regarding material selection, device design, and fabrication methodology. The inductors achieved here present the best performance to date for an integrated magnetic core inductor at high frequencies with a 1 nH inductance and peak quality factor of 4 at $\sim$3 GHz. Such compact inductors show potential for efficiently meeting the need of mobile electronics in the future.
\end{abstract}

\section{Introduction}
In the past couple decades, tremendous effort has been put forth towards incorporating thin-film magnetic cores into integrated passive circuit elements, including inductors and transformers \cite{Dhagat2004, Mehas1999, Sullivan2003, Wang2007, Prabhakaran2004, Meere2011, Sturcken2013, Lee2008,Ikeda2003, Gardner2009, Gardner2007, Salvia2005, Gao2014, Yun2004, Mullenix2013}, for a variety of applications. The traditional operating frequency and flux-amplification properties of integrated magnetic films have made them ideal for power management applications up to now \cite{Dhagat2004, Mehas1999, Sullivan2003, Wang2007, Prabhakaran2004, Meere2011, Sturcken2013, Lee2008, Mullenix2013, Rhen2008,Leary2012, Shen2012}. However, the cost of depositing such films thick enough for power electronics devices has limited their usage in integrated systems to date. More recently, as mobile electronics have begun to demand higher inductances and quality factors in smaller form factors than previously achievable by spiral air-core inductors in that frequency range \cite{ITRS}, magnetic cores are being redesigned to compete to meet the new need \cite{Gardner2007, Salvia2005, Gao2014}.

\begin{figure}[]
\centering
\includegraphics[width=3.5in]{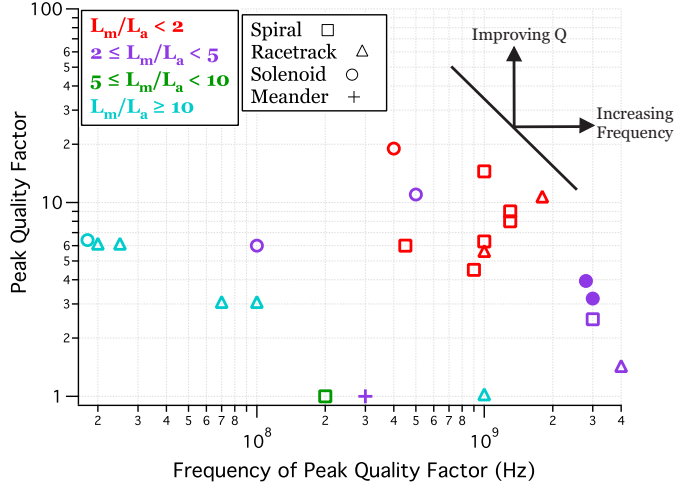}
\caption{A summary of published magnetic-core inductor and transformer device performances, comparing quality factor and peak frequency \cite{Yamaguchi2000_2,Yamaguchi2005,Yamaguchi2000,Yamaguchi2000_3,Xu2011,Arai1991,Xu2011_2, Yamaguchi2001,Lee2008,Ikeda2003, Gardner2009, Gardner2007, Salvia2005, Gao2014,Mullenix2013}. The color of each marker indicates the inductance enhancement, $L_m/L_a$, of the device compared to its air-core equivalent. The inductance enhancement was evaluated within the frequency band where the inductance is flat. Of the four inductors clustered at high frequencies, two belong to this work, while the other two belong to \cite{Xu2011_2}. The solid circles mark this work, embodying a combined improvement towards higher Q factor and operating frequency. }
\label{Review}
\end{figure}

While the high permeability of magnetic materials increases the DC value of inductance, the challenge often faced is to extend that permeability and corresponding inductance enhancement to higher frequencies of use to mobile applications (namely in the 1-5 GHz range). At the intrinsic ferromagnetic resonance (FMR) frequency of magnetic materials (typically between 1-2 GHz for large blanket films), the relative permeability drops to unity and magnetic loss tangent peaks, therefore making the inductance enhancement due to the material negligible and, instead, the losses due to it dominant. Nevertheless, through various means of deposition and patterning the frequency response of the magnetic permeability can be improved and extended over a larger bandwidth. Figure \ref{Review} summarizes the performance of various high frequency magnetic-core inductors from literature.

This work provides a detailed examination of the optimization choices involved in maximizing the inductance enhancement and bandwidth, while simultaneously attempting to maximize the quality factor of magnetic-core inductors. This paper discusses the most crucial considerations regarding material selection and fabrication process as well as magnetic-core and device design. Such optimizations were then applied, yielding a fabricated device with the highest combined performance for an integrated magnetic-core inductor at high frequencies; the device offers a magnetic enhancement over air core extending beyond 6 GHz, a low frequency inductance of 1nH, and a peak quality factor of 4 at approximately 3 GHz.

\section{Design Considerations}
\subsection{Inductance and Bandwidth}
	High frequency performance of magnetic inductors depends heavily upon the magnetic-core material and design. High permeability magnetic materials provide greater inductance enhancements over their air-core equivalents, but simultaneously suffer from higher eddy currents and lower FMR frequencies. Elimination of eddy currents by lamination - breaking up the conductive magnetic core with insulating spacers to limit the flow of eddy currents - is one component of improving the high frequency behavior, however the largest contribution to the improved frequency response of the inductors comes from increasing the FMR frequency and thereby reducing magnetic losses in the frequency range of interest. Extending the bandwidth of magnetic materials by increasing the FMR frequency can be done in a couple of ways involving using the shape of the magnetic material to change its uniaxial anisotropy and resultant permeability spectrum. Uniaxial anisotropy keeps the magnetization of the inductor core in the plane of the film with the AC permeability measured perpendicular to the magnetization direction corresponding to the rotation of magnetization in a small signal applied magnetic field.

\begin{figure}[!t]
    \centering
    \begin{subfigure}[t]{0.45\textwidth}
        \centering
        \includegraphics[width=3in]{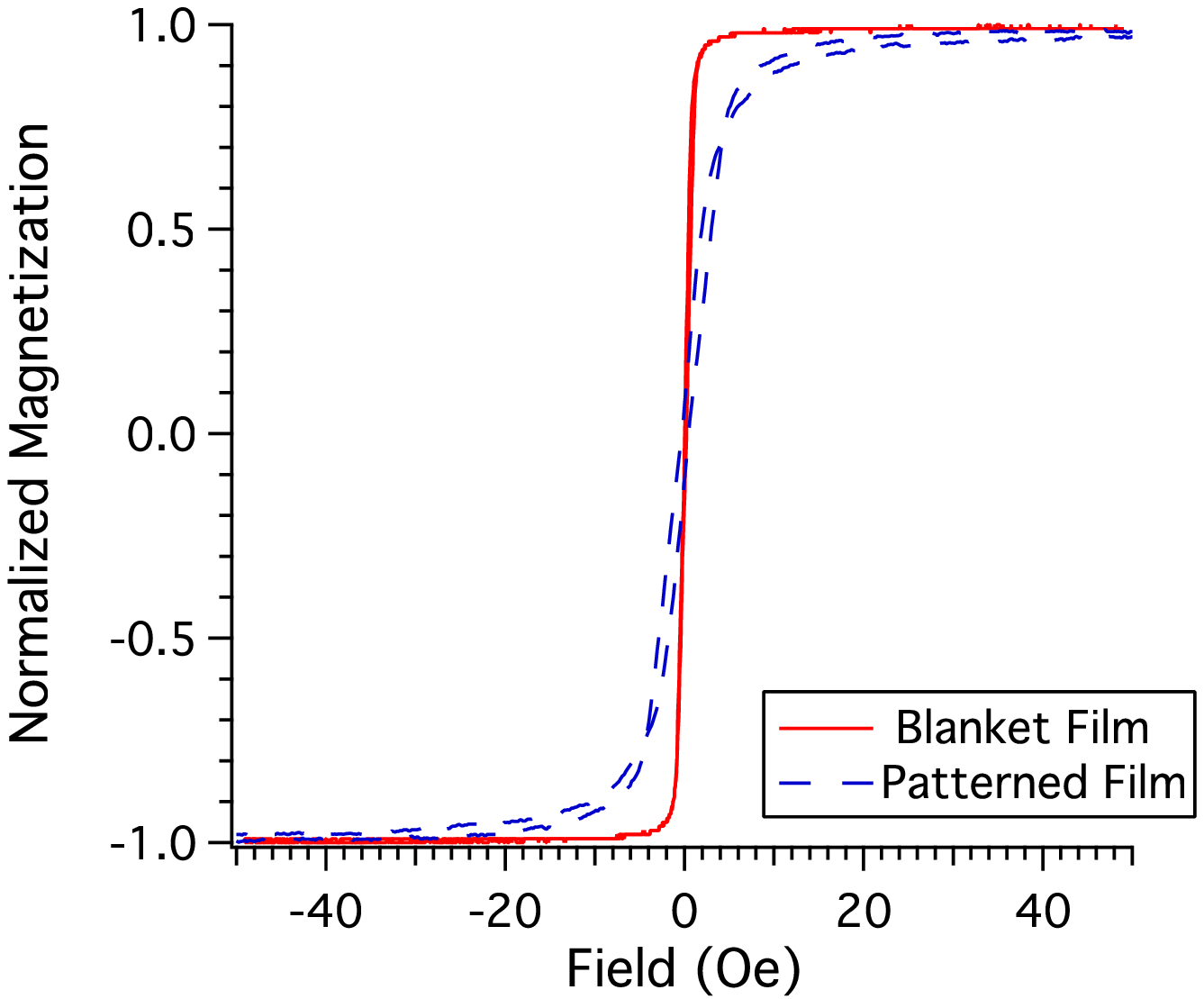}
        \caption{Along Length (Easy Axis)}
	\label{EAloops}
    \end{subfigure}
\hfill
    \begin{subfigure}[t]{0.45\textwidth}
        \centering
        \includegraphics[width=3in]{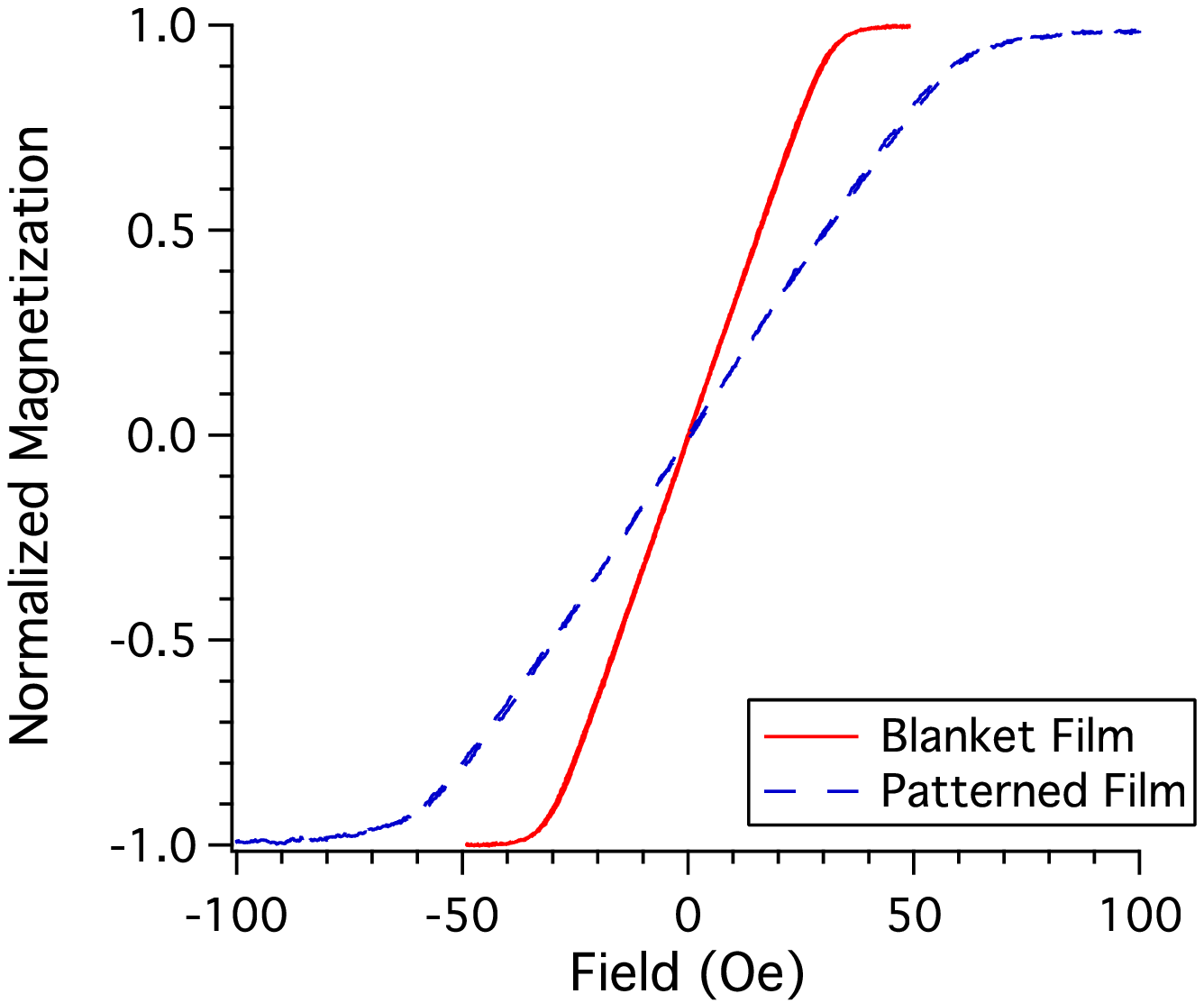}
        \caption{Along Width (Hard Axis)}
	\label{HAloops}
    \end{subfigure}
    \caption{Magnetization hysteresis loops along the a) length and b) width directions of the magnetic core. The width direction is orthogonal to the uniaxial magnetic anisotropy and gives insight into the permeability for small signal magnetic fields.}
\label{BHloops}
\end{figure}

The first method of increasing the FMR frequency is to pattern the dimensions of the core so as to maximize the length of the magnetic core parallel to its in-plane uniaxial magnetization. This practice has been shown to increase the FMR frequency while consequently also reducing the permeability as a fundamental trade-off \cite{Perrin1997,Kim2007, Vroubel2004,Ikeda2005, Walser1998, Webb1991, Yamaguchi2000, Yamaguchi2005, Yamaguchi2000_2}. The second method builds on this and utilizes ultra-thin laminations in the thickness direction to also increase the FMR frequency \cite{El-Ghazaly2015}. Both of these methods were combined in the design of the magnetic core inductors described in this paper. The core was patterned into a 250 $\mu$m wide x 1000 $\mu$m long rectangular bar with the magnetization along the length direction. Figure \ref{BHloops} shows the overlapped magnetization hysteresis curves for the blanket and patterned films, indicating an increase in anisotropy field and a decrease in permeability for the patterned film. The anisotropy, marked by the field required for saturation, was seen to increase by a factor of 2$\times$. 

\begin{figure}[!t]
\centering
\includegraphics[width=3.5in]{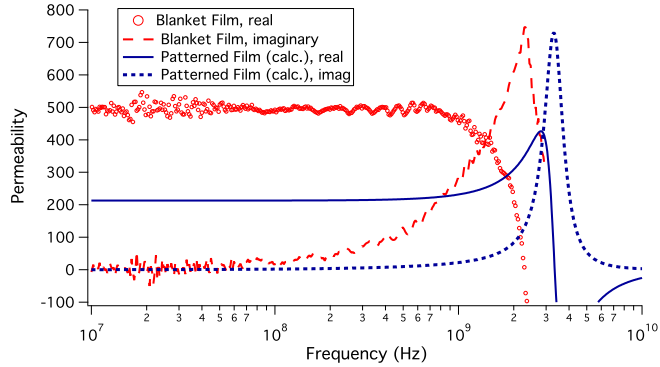}
\caption{Permeability spectra for both the blanket and patterned magnetic films. Data for the blanket film permeability spectrum is obtained through measurement, while that of the patterned film is calculated using eqn. \ref{analyticalPermeability} and approximate demagnetization factors obtained through simulation ($N_z = 0.9952$, $N_x = 0.00384$, and $N_y = 0.00096$). The FMR frequency of the blanket film is $\sim$2.2 GHz, while that of the patterned film is $\sim$3.3 GHz (above the 3 GHz limit of the measurement setup).}
\label{Permeability}
\end{figure}

A comparison of the permeability spectra of the blanket (measured) and patterned film (calculated) in Fig. \ref{Permeability} confirms that there is a corresponding decrease in the low frequency permeability value. The shape-dependent permeability spectrum of the magnetic core was calculated according to eqn. \ref{analyticalPermeability}, where the material parameters were measured and where the demagnetization factors were simulated using the same method described in \cite{El-Ghazaly2015}.
\begin{equation}
\label{analyticalPermeability}
\mu_r = 1+\frac{(\omega_k+\omega_{zy}+i\alpha\omega)\omega_m}{(\omega_k+\omega_{xy}+i\alpha\omega)(\omega_k+\omega_{zy}+i\alpha\omega)-\omega^2}
\end{equation}
where
\begin{equation}
\label{wk}
\omega_k = \gamma \mu_0 H_k
\end{equation}
\begin{equation}
\label{wm}
\omega_m = \gamma \mu_0 M_s
\end{equation}
\begin{equation}
\label{wNzy}
\omega_{zy} = \gamma \mu_0 (N_z-N_y) M_s
\end{equation}
\begin{equation}
\label{wNxy}
\omega_{xy} = \gamma \mu_0 (N_x-N_y) M_s
\end{equation}
$\gamma$ is the gyromagnetic ratio, $H_k$ is the magnetic anisotropy field, $M_s$ is the saturation magnetization, and $N_z$, $N_x$, and $N_y$ are the demagnetization factors along the out-of-plane dimension, width dimension, and length dimension (easy-axis), respectively. In the measured film spectra, the FMR peak in the blanket film appears suppressed, likely as a result of large eddy currents or interlaminate capacitance \cite{Webb1991}.

The FMR frequency can be specifically calculated by the following Kittel equation:
\begin{equation}
\label{kittel}
\omega_{_{FMR}} = \gamma \mu_0\sqrt{\left [H_k+(N_z-N_y) M_s \right ]\left [H_k+(N_x-N_y) M_s \right ]}
\end{equation}
Nevertheless, by looking at the imaginary permeability peak location, which represents the FMR frequency and the point at which magnetic losses begin to dominate, a direct comparison can be made of the bandwidth of the magnetic cores. Therefore, patterning of the magnetic core effectively increased its bandwidth by more than 1 GHz.

\subsection{Quality Factor}
Increasing the bandwidth is accompanied by the consequential decrease in the permeability, as seen in Fig. \ref{Permeability}. Therefore, although pushing the magnetic losses out to higher frequencies potentially improves the quality factor (Q), the inductance decrease with permeability somewhat negates the effect. In order to compensate for the reduction in inductance while also minimizing the resistance of the inductor, the structure was extensively simulated for both maximum inductance and maximum Q factor.

The greatest two design parameters contributing to increasing the inductance (with least effect on the resistance) are reducing the width of the windings and reducing the spacing between windings. Both of these reductions work to increase the efficiency of flux linkage from one winding to the next without increasing the length the solenoid coil. The final width and spacing selected were each 20 $\mu$m. Although reducing the width of the windings increases inductance, it also increases the resistance of the coil. Therefore, after optimizing for inductance, the resistance had to be optimized to improve the Q. Increasing the thickness of the coil can accomplish this; while it has minimal effect on the inductance, it increases the Q because it reduces the coil resistance. Due to fabrication constraints, however, the coil thickness in this work was limited to 3 $\mu$m thick copper windings and the magnetic core thickness was limited to 1 $\mu$m thick CoFeB film. Higher Q and higher inductance would be expected by increasing the thicknesses of each, respectively. Despite the fabrication limitations, this work demonstrates a major push forward for magnetic inductors operating at high frequencies.

\section{Materials and Fabrication}
\begin{figure}[!t]
    \centering
    \begin{subfigure}[t]{0.45\textwidth}
        \centering
        \includegraphics[width=2in]{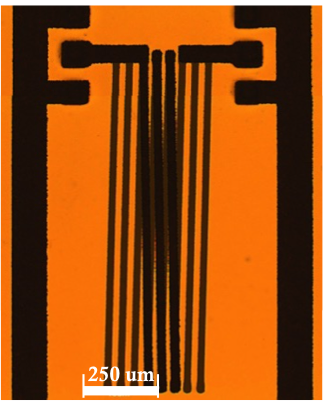}
        \caption{Air Core Inductor}
	\label{AirInductor}
    \end{subfigure}
\hfill
    \begin{subfigure}[t]{0.45\textwidth}
        \centering
        \includegraphics[width=2in]{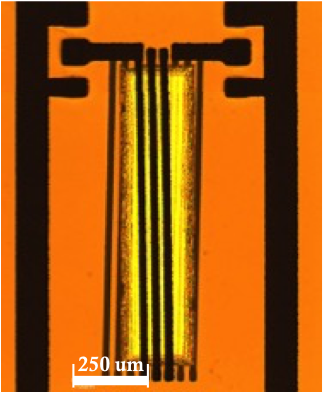}
        \caption{Magnetic Core Inductor}
	\label{MagInductor}
    \end{subfigure}
    \caption{Optical images of the fabricated a) air-core and b) magnetic-core versions of the same inductor design.}
\label{Inductors}
\end{figure}

The solenoid inductors shown in Fig. \ref{Inductors} were fabricated to demonstrate a direct comparison between air-core (left) and magnetic-core (right) inductors of the same design to evaluate the contribution due to the magnetic film. Fabrication of the inductors involved six major phases: isolation, bottom inductor windings, planarization and insulation, magnetic core, planarization and insulation, and vias and top inductor windings. In the first phase, 40 $\mu$m of the commercial polymer SU-8 2015 was formed on the surface of a 4 in silicon wafer to provide sufficient isolation between the radio frequency (RF) inductor and potential substrate parasitics. Bottom inductor windings, top inductor windings, and vias were deposited using a through-mold electroplating process: a Ti/Cu seed layer was first evaporated onto the wafer, photoresist patterns provided walls for a subsequent electroplating process to build ~3 $\mu$m thick Cu windings, and finally, the photoresist was removed and the non-plated seed layer areas were etched away. A two-layer SU-8 2002 polymer process was developed to first fill the gaps between windings for planarization and then provide an additional 1.5 $\mu$m insulation layer between the windings and the magnetic layer.

The material CoFeB was selected for the magnetic core based on its previously demonstrated excellent high frequency properties \cite{Chen2000,Lebedev2012}. The magnetic core was laminated to break up the conductive magnetic core with insulating spacers that further improve broadband performance by reducing high frequency losses due to eddy current circulation in the thickness direction. The magnetic core was deposited in a single vacuum chamber under an applied magnetic field used to induce a uniaxial magnetic anisotropy along the length of the inductor. RF sputter deposition alternated between 63 nm Co\textsubscript{43}Fe\textsubscript{43}B\textsubscript{14} (At\%) and 6 nm SiO\textsubscript{2} to construct a 16x(CoFeB/SiO\textsubscript{2}) laminated magnetic stack with total thickness of approximately 1 $\mu$m. The stack was then patterned using optical photolithography and etched in a 1:1:2 ratio of HF:HNO\textsubscript{3}:H\textsubscript{2}O. Proper alignment between the external magnetic field during deposition and the subsequent lithographic patterning of the cores is critical for ensuring the ideal, linear, easy-axis domains in the magnetic core that provide broadband constant inductance \cite{El-Ghazaly2013}.

\section{Results and Discussion}
\begin{figure}[!t]
    \centering
    \begin{subfigure}[t]{0.5\textwidth}
        \centering
        \includegraphics[width=3in]{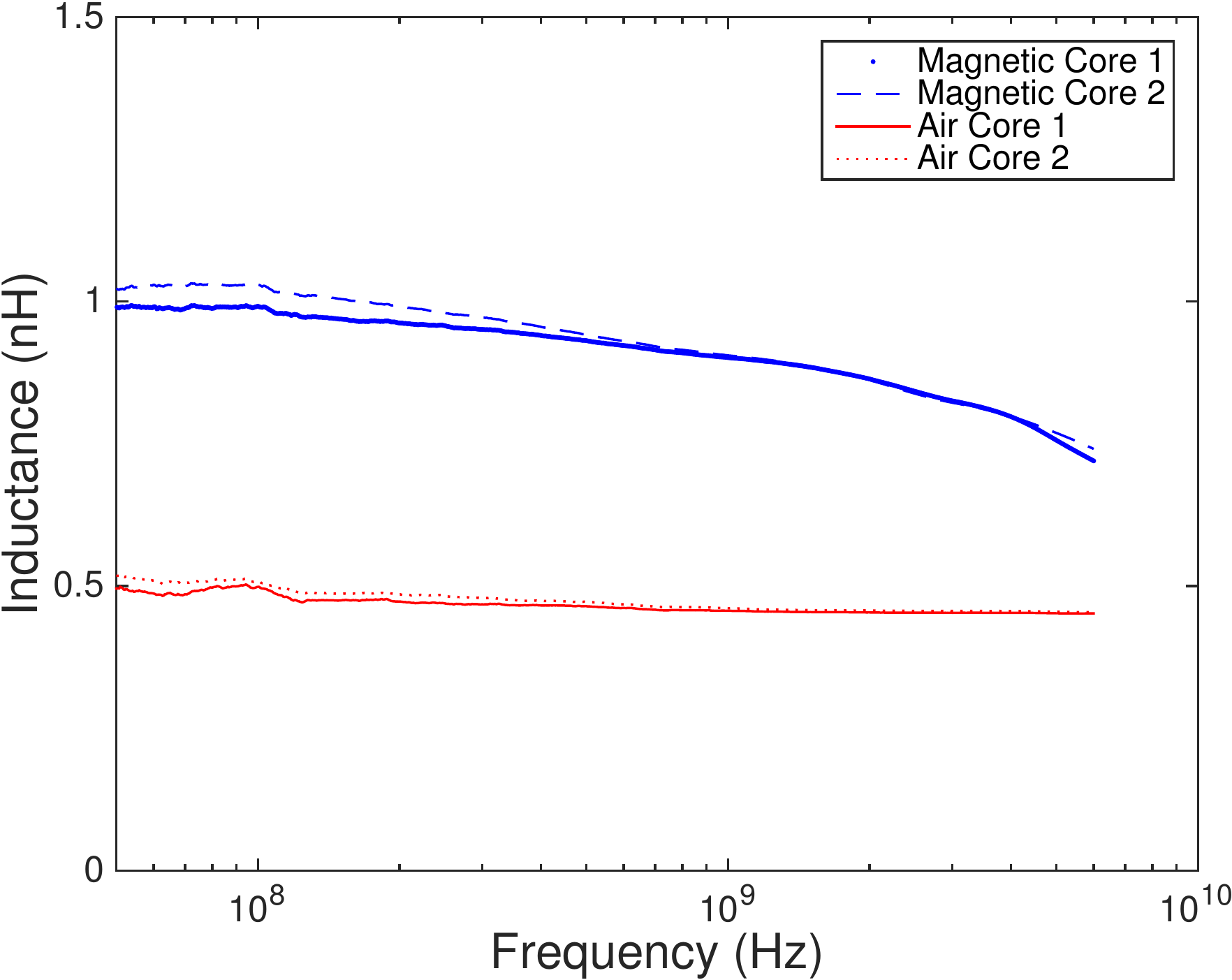}
        \caption{Inductance}
	\label{L}
    \end{subfigure}
\hfill
    \begin{subfigure}[t]{0.5\textwidth}
        \centering
        \includegraphics[width=3in]{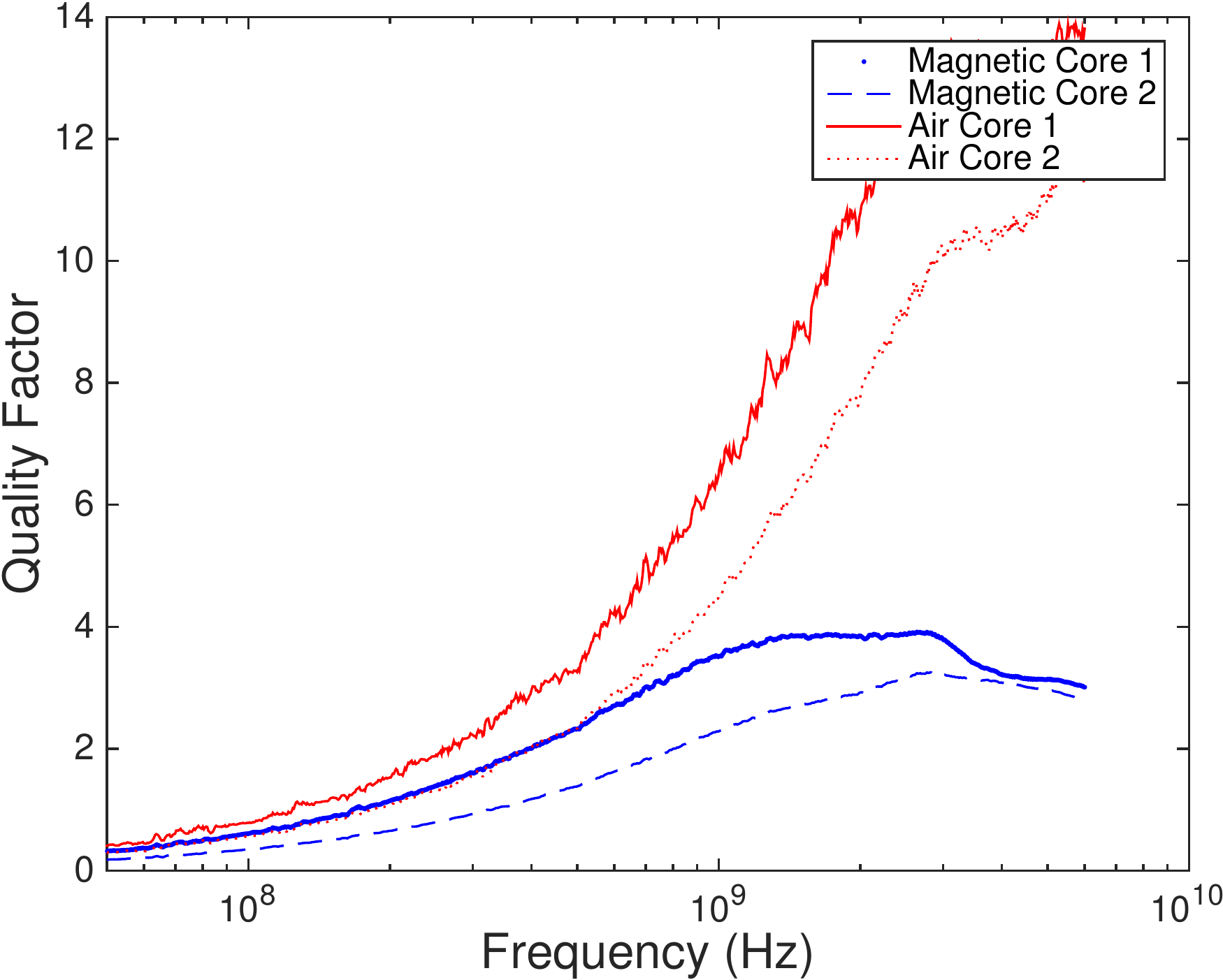}
        \caption{Quality Factor}
	\label{Q}
    \end{subfigure}
\hfill
    \begin{subfigure}[t]{0.5\textwidth}
        \centering
        \includegraphics[width=3in]{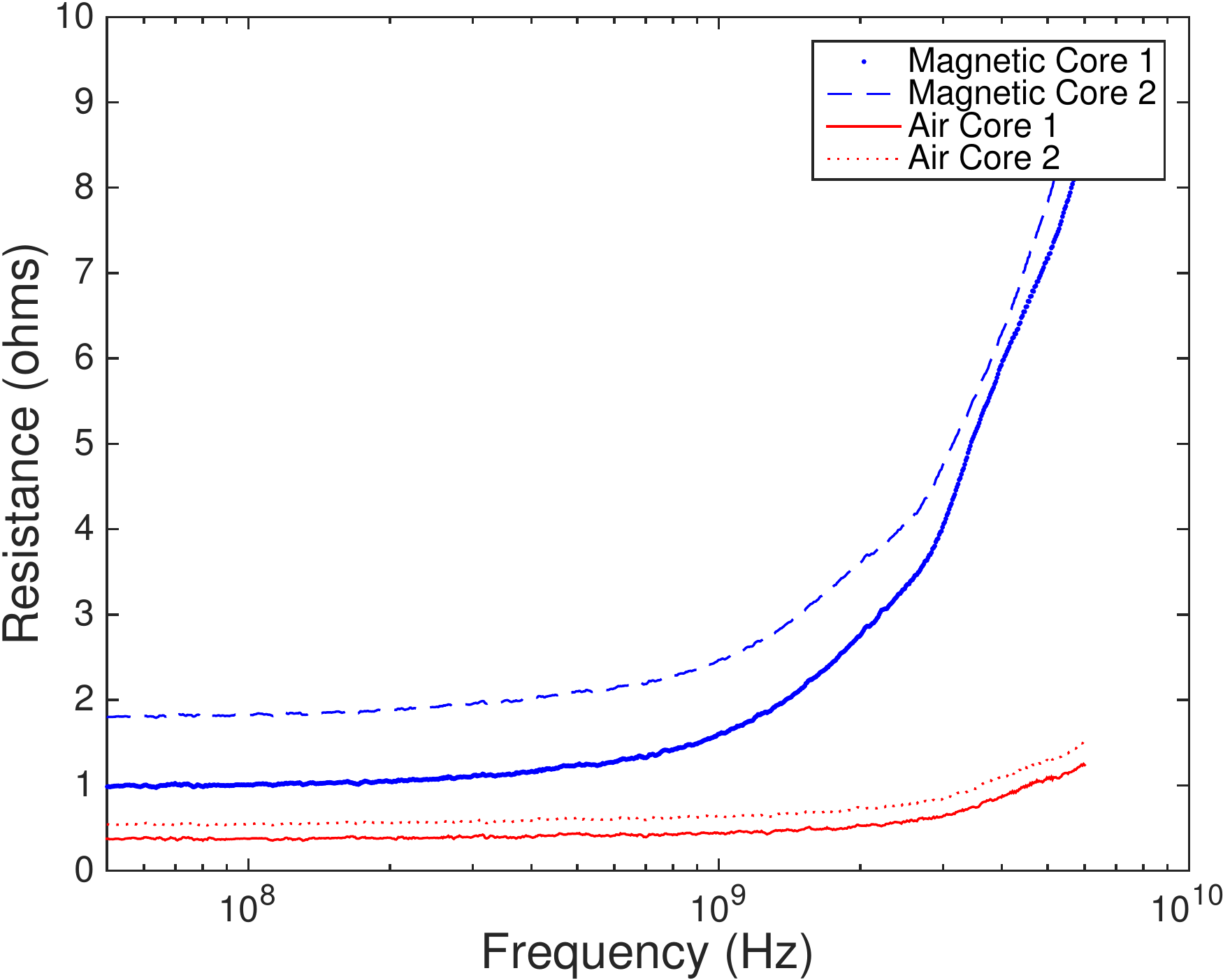}
        \caption{Resistance}
	\label{R}
    \end{subfigure}
    \caption{Measured a) inductance, b) quality factor, and c) resistance results for both the air-core and a couple of magnetic-core inductors of the same design. The magnetic-core shows a 100\% increase of inductance from the 0.5 nH of the air-core to 1 nH.}
\label{Results}
\end{figure}

To directly evaluate the performance enhancement of the inductor due to the magnetic material, similar inductors both with and without the magnetic core were measured. The inductance, quality factor (Q), and resistance results are presented in Fig. \ref{Results}. Incorporation of the magnetic core is seen to double the inductance of the air-core inductor, providing approximately 1 nH inductance into the GHz range. Decreasing inductance at higher frequencies is likely due to ferromagnetic resonance, as well as a combination of misalignment of the applied magnetic field during deposition of the cores \cite{El-Ghazaly2013}, parasitic capacitances, and eddy currents along the length of the core. Evidence of misalignment of the magnetic anisotropy can be seen from the hysteresis loops in Fig. \ref{EAloops}.

A peak quality factor of 4 was achieved at approximately 3 GHz, representing the best performance of a high-frequency magnetic-core inductor to date. Increase in frequency is attributed to the increase in bandwidth of the magnetic core by pushing the FMR frequency higher. This understanding is supported by the parallel trends in both the high frequency resistance results and the typical imaginary permeability spectrum (representing magnetic losses). This direct correspondence indicates that the main losses contributing to the increase in resistance at high frequencies and the decrease in quality factor are due to the magnetic core. Substrate effects are negligible due to the very thick, 40 $\mu$m, isolation layer separating the inductor from the Si substrate.

For best comparison, the DC resistances of the inductors should be identical. However, since the inductors were selected from different locations on the wafer some wafer-level process variations (coil thickness, etch-back, via-opening, etc.), the low-frequency resistances of the inductors differ slightly from each other, leading to undesired reductions in Q for the inductors with higher resistance. Nevertheless, as expected, the inductor with higher inductance and lower DC resistance yielded the highest quality factor. Compared to previously published air-core \cite{Lopez-Villegas1998,Meyer2010,Raieszadeh2005} and magnetic-core \cite{Lee2008,Ikeda2003, Gardner2009, Gardner2007, Salvia2005, Gao2014, Yamaguchi2000, Yamaguchi2005, Yamaguchi2000_2,Frommberger2005,Wang2016} inductors, the results suggest a very competitive new advancement of magnetic inductors for RF power amplifiers and other applications.

While previous work has demonstrated an increase in inductor frequency due to similar patterning for higher FMR frequencies, either the inductance enhancements of those designs were very low ($<$1.2$\times$) \cite{Yamaguchi2000, Yamaguchi2005, Yamaguchi2000_2} or the Q-factors were low ($\le$2.5) \cite{Xu2011_2}. The inductors shown here are the first to demonstrate overall optimization of inductance enhancement, quality factor, and operating frequency, simultaneously. Of the previously reported integrated magnetic inductors that had similar inductance or inductance enhancement values, much lower peak quality factor frequencies were observed \cite{Ikeda2003,Gardner2007, Salvia2005, Gao2014}. In some cases, the value of the peak Q-factor was similar \cite{Ikeda2003,Gardner2007}, but in other cases, the Q was in the range of 10-20 \cite{Salvia2005, Gao2014}. The slightly lower quality factor in the current design can be attributed to a number of factors including the increased DC inductor winding resistance due to narrower and thinner copper lines (20 $\mu$m wide and 3 $\mu$m thick in this work compared to ~150 $\mu$m wide and 4.5 $\mu$m thick in \cite{Gao2014}) and other process variations. Narrower inductor windings, however, result in better inductive coupling, thereby increasing the inductance and compensating for the decrease in permeability accompanying the narrow patterning of the magnetic inductor. Furthermore, with larger copper thickness, the quality factor is expected to increase to comparable or higher values than those reported in other works. These results in addition to simulations confirm that magnetic-core inductors can, indeed, be used to satisfy the requirements of future RF and mobile applications.

\section{Conclusion}
The methodology for design and fabrication of high-frequency magnetic-core inductors was presented. Furthermore, integrated magnetic inductors have been demonstrated with increased frequency bandwidth to meet the need for GHz-range power amplifier applications. The thin-film magnetic core inductors have an inductance of 1 nH, embodying a $2\times$ inductance enhancement over its air-core counterpart at low frequencies and a $1.6\times$ enhancement at the frequency of the peak quality factor. The 3 GHz frequency and peak quality factor of 4 are the highest reported combined performance to date for a magnetic core inductor. Extension of the operating frequency of the inductor is attributed to the use of patterning and laminations to increase shape anisotropy in the magnetic material and increase its ferromagnetic resonance frequency, reducing the magnetic losses within the desired operating frequency range. With continued design optimization of the inductor windings and magnetic core, it is expected that the unmet need for high-Q, high-frequency inductors may soon be met with magnetic materials.

\section*{Acknowledgment}
Research for this project was conducted with government support under FA9550-11-C-0028 and awarded by the Department of Defense, Air Force Office of Scientific Research, National Defense Science and Engineering Graduate (NDSEG) Fellowship, 32 CFR 168a. Part of this work was performed using the Stanford Nanofabrication Facility (SNF) and at the Stanford Nano Shared Facilities (SNSF) at Stanford University.


\begin{thebibliography}{10}
\providecommand{\url}[1]{#1}
\csname url@samestyle\endcsname
\providecommand{\newblock}{\relax}
\providecommand{\bibinfo}[2]{#2}
\providecommand{\BIBentrySTDinterwordspacing}{\spaceskip=0pt\relax}
\providecommand{\BIBentryALTinterwordstretchfactor}{4}
\providecommand{\BIBentryALTinterwordspacing}{\spaceskip=\fontdimen2\font plus
\BIBentryALTinterwordstretchfactor\fontdimen3\font minus
  \fontdimen4\font\relax}
\providecommand{\BIBforeignlanguage}[2]{{%
\expandafter\ifx\csname l@#1\endcsname\relax
\typeout{** WARNING: IEEEtran.bst: No hyphenation pattern has been}%
\typeout{** loaded for the language `#1'. Using the pattern for}%
\typeout{** the default language instead.}%
\else
\language=\csname l@#1\endcsname
\fi
#2}}
\providecommand{\BIBdecl}{\relax}
\BIBdecl

\bibitem{Dhagat2004}
P.~Dhagat, S.~Prabhakaran, and C.~R. Sullivan, ``{Comparison of Magnetic
  Materials for V-Groove Inductors in Optimized High-Frequency},'' \emph{IEEE
  Transactions on Magnetics}, vol.~40, no.~4, pp. 2008--2010, 2004.

\bibitem{Mehas1999}
G.~Mehas, K.~Coonley, and C.~Sullivan, ``{Design of microfabricated inductors
  for microprocessor power delivery},'' \emph{Applied Power Electronics
  Conference}, pp. 1181--1187, 1999.

\bibitem{Sullivan2003}
C.~R. Sullivan, S.~Prabhakaran, P.~Dhagat, and Y.~Sun, ``{Thin-Film Inductor
  Designs and Materials for High-Current Low-Voltage Power},''
  \emph{Transactions of the Magnetics Society of Japan}, vol.~3, no.~4, pp.
  126--8, 2003.

\bibitem{Wang2007}
N.~Wang, T.~O'Donnell, S.~Roy, P.~McCloskey, O', and C.~Mathuna,
  ``{Micro-inductors integrated on silicon for power supply on chip},''
  \emph{Journal of Magnetism and Magnetic Materials}, vol. 316, 2007.

\bibitem{Prabhakaran2004}
S.~Prabhakaran, C.~R. Sullivan, T.~O'Donnell, M.~Brunet, and S.~Roy,
  ``{Microfabricated coupled inductors for DC-DC converters for microprocessor
  power delivery},'' \emph{PESC Record - IEEE Annual Power Electronics
  Specialists Conference}, vol.~6, pp. 4467--4472, 2004.

\bibitem{Meere2011}
R.~Meere, N.~Wang, T.~O'Donnell, S.~Kulkarni, S.~Roy, and S.~C. O'Mathuna,
  ``{Magnetic-core and air-core inductors on silicon: A performance comparison
  up to 100 MHz},'' \emph{IEEE Transactions on Magnetics}, vol.~47, no.~10, pp.
  4429--4432, 2011.

\bibitem{Sturcken2013}
N.~Sturcken, E.~J.~O. Sullivan, N.~Wang, P.~Herget, B.~C. Webb, L.~T. Romankiw,
  M.~Petracca, R.~Davies, R.~E. Fontana, G.~M. Decad, I.~Johnkymissis, A.~V.
  Peterchev, L.~P. Carloni, W.~J. Gallagher, and K.~L. Shepard, ``{A 2.5D
  Integrated Voltage Regulator Using Coupled-Magnetic-Core Inductors on Silicon
  Interposer},'' \emph{IEEE Journal of Solid-State Circuits}, vol.~48, no.~1,
  pp. 244--254, 2013.

\bibitem{Lee2008}
D.~W. Lee, K.-p. Hwang, and S.~X. Wang, ``{Fabrication and Analysis of
  High-Performance Integrated Solenoid Inductor With Magnetic Core},''
  \emph{IEEE Transactions on Magnetics}, vol.~44, no.~11, pp. 4089--4095, 2008.

\bibitem{Ikeda2003}
K.~Ikeda, K.~Kobayashi, K.~Ohta, R.~Kondo, T.~Suzuki, and M.~Fujimoto,
  ``{Thin-film inductor for gigahertz band with CoFeSiO-SiO/sub 2/2 multilayer
  granular films and its application for power amplifier module},'' \emph{IEEE
  Transactions on Magnetics}, vol.~39, no.~5, pp. 3057--3061, sep 2003.

\bibitem{Gardner2009}
D.~Gardner, G.~Schrom, F.~Paillet, B.~Jamieson, T.~Karnik, and S.~Borkar,
  ``{Review of On-Chip Inductor Structures With Magnetic Films},'' \emph{IEEE
  Transactions on Magnetics}, vol.~45, no.~10, pp. 4760--4766, oct 2009.

\bibitem{Gardner2007}
D.~S. Gardner, G.~Schrom, P.~Hazucha, F.~Paillet, T.~Karnik, and S.~Borkar,
  ``{Integrated on-chip inductors with magnetic films},'' \emph{IEEE
  Transactions on Magnetics}, vol.~43, no.~6, pp. 2615--2617, 2007.

\bibitem{Salvia2005}
J.~Salvia, J.~Bain, and C.~Yue, ``{Tunable on-chip inductors up to 5 GHz using
  patterned permalloy laminations},'' \emph{IEEE International Electron Devices
  Meeting 2005}, vol.~00, no.~1, pp. 963--966, 2005.

\bibitem{Gao2014}
Y.~Gao, S.~Z. Zardareh, X.~Yang, T.~X. Nan, Z.~Y. Zhou, M.~Onabajo, M.~Liu,
  A.~Aronow, K.~Mahalingam, B.~M. Howe, G.~J. Brown, and N.~X. Sun,
  ``{Significantly enhanced inductance and quality factor of GHz integrated
  magnetic solenoid inductors with FeGaB/Al2O3 Multilayer Films},'' \emph{IEEE
  Transactions on Electron Devices}, vol.~61, no.~5, pp. 1470--1476, 2014.

\bibitem{Yun2004}
E.~Yun, M.~Jung, C.~Cheon, and H.~G. Nam, ``{Microfabrication and
  characteristics of low-power high-performance magnetic thin-film
  transformers},'' \emph{IEEE Transactions on Magnetic}, vol.~40, no.~1, pp.
  65--70, 2004.

\bibitem{Mullenix2013}
J.~Mullenix, A.~El-Ghazaly, and S.~X. Wang, ``{Integrated Transformers With
  Sputtered Laminated Magnetic Core},'' \emph{IEEE Transactions on Magnetics},
  vol.~49, no.~7, pp. 4021--4027, 2013.

\bibitem{Rhen2008}
F.~M. Rhen, P.~McCloskey, T.~O'Donnell, and S.~Roy, ``{High-frequency
  permeability of electroplated CoNiFe and CoNiFe-C alloys},'' \emph{Journal of
  Magnetism and Magnetic Materials}, vol. 320, no.~20, pp. e819--e822, oct
  2008.

\bibitem{Leary2012}
A.~M. Leary, P.~R. Ohodnicki, and M.~E. Mchenry, ``{Soft magnetic materials in
  high-frequency, high-power conversion applications},'' \emph{Journal of
  Minerals, Metals, and Materials}, vol.~64, no.~7, pp. 772--781, 2012.

\bibitem{Shen2012}
S.~Shen, P.~R. Ohodnicki, S.~J. Kernion, A.~M. Leary, V.~Keylin, J.~F. Huth,
  and M.~E. Mchenry, ``{Nanocomposite Alloy Design for High Frequency Power
  Conversion Applications},'' \emph{Energy Technology 2012: Carbon Dioxide
  Management and Other Technologies}, pp. 275--282, 2012.

\bibitem{ITRS}
``{International Technology Roadmap for Semiconductors},'' Tech. Rep., 2011.

\bibitem{Yamaguchi2000_2}
M.~Yamaguchi, M.~Baba, K.~Suezawa, T.~Moizumi, K.~I. Arai, Y.~Shimada, A.~Haga,
  S.~Tanabe, and K.~Ito, ``{Magnetic RF integrated thin-film inductors},'' in
  \emph{Microwave Symposium Digest. 2000 IEEE MTT-S International}, no.~c,
  2000, pp. 205--208.

\bibitem{Yamaguchi2005}
M.~Yamaguchi, S.~Bae, K.~H. Kim, K.~Tan, T.~Kusumi, and K.~Yamakawa,
  ``{Ferromagnetic RF integrated inductor with closed magnetic circuit
  structure},'' in \emph{IEEE MTT-S International Microwave Symposium Digest},
  2005, pp. 351--354.

\bibitem{Yamaguchi2000}
M.~Yamaguchi, K.~Suezawa, Y.~Takahashi, K.~I. Arai, S.~Kikuchi, Y.~Shimada,
  S.~Tanabe, and K.~Ito, ``{Magnetic thin-film inductors for RF-integrated
  circuits},'' \emph{Journal of Magnetism and Magnetic Materials}, vol. 215,
  pp. 807--810, 2000.

\bibitem{Yamaguchi2000_3}
M.~Yamaguchi, K.~Suezawa, M.~Baba, K.~Arai, Y.~Shimada, S.~Tanabe, and K.~Itoh,
  ``{Application of bi-directional thin-film micro wire array to RF integrated
  spiral inductors},'' \emph{IEEE Transactions on Magnetics}, vol.~36, no.~5,
  pp. 3514--3517, 2000.

\bibitem{Xu2011}
W.~Xu, S.~Sinha, T.~Dastagir, H.~Wu, B.~Bakkaloglu, D.~S. Gardner, Y.~Cao, and
  H.~Yu, ``{Performance enhancement of on-chip inductors with permalloy
  magnetic rings},'' \emph{IEEE Electron Device Letters}, vol.~32, no.~1, pp.
  69--71, 2011.

\bibitem{Arai1991}
K.~I. Arai, M.~Yamaguchi, H.~Ohzeki, and M.~Matsumoto, ``{Application of YIG
  film to thin film inductors},'' \emph{IEEE Transactions on Magnetics},
  vol.~27, no.~6, pp. 5337--5339, 1991.

\bibitem{Xu2011_2}
W.~Xu, H.~Wu, D.~S. Gardner, S.~Sinha, T.~Dastagir, B.~Bakkaloglu, Y.~Cao, and
  H.~Yu, ``{Sub-100 $\mu$m scale on-chip inductors with CoZrTa for GHz
  applications},'' \emph{Journal of Applied Physics}, vol. 109, no.~7, p.
  07A316, 2011.

\bibitem{Yamaguchi2001}
{Masahiro Yamaguchi}, M.~Baba, and K.-I. Arai, ``{Sandwich-Type Ferromagnetic
  RF Integrated Inductor},'' \emph{IEEE Transactions on Microwave Theory and
  Techniques}, vol.~49, no.~12, pp. 2331--2335, 2001.

\bibitem{Perrin1997}
G.~Perrin, J.~C. Peuzin, and O.~Acher, ``{Control of the resonance frequency of
  soft ferromagnetic amorphous thin films by strip patterning},'' \emph{Journal
  of Applied Physics}, vol.~81, no.~8, p. 5166, 1997.

\bibitem{Kim2007}
K.~Kim, M.~Yamaguchi, and J.~Kim, ``{Ferromagnetic resonance behaviors of
  integrated CoPdAlO magnetic film on coplanar waveguide},'' \emph{Journal of
  the Korean Physical Society}, vol.~51, no.~6, pp. 2026--2030, 2007.

\bibitem{Vroubel2004}
M.~Vroubel, Y.~Zhuang, B.~Rejaei, J.~Burghartz, a.M. Crawford, and S.~Wang,
  ``{Calculation of Shape Anisotropy for Micropatterned Thin Fe–Ni Films for
  On-Chip RF Applications},'' \emph{IEEE Transactions on Magnetics}, vol.~40,
  no.~4, pp. 2835--2837, Jul. 2004.

\bibitem{Ikeda2005}
S.~Ikeda, K.~H. Kim, and M.~Yamaguchi, ``{Slit-patterned CoNbZr/Nb/CoNbZr
  magnetic film for rf noise suppressor},'' \emph{Journal of Applied Physics},
  vol.~97, no.~10, p. 10F912, 2005.

\bibitem{Walser1998}
R.~M. Walser, W.~Win, and P.~M. Valanju, ``{Shape-Optimized Ferromagnetic
  Particles with Maximum Theoretical Microwave Susceptibility},'' \emph{IEEE
  Transactions on Magnetics}, vol.~34, no.~4, pp. 1390--1392, 1998.

\bibitem{Webb1991}
B.~Webb and M.~Re, ``{High-frequency permeability of laminated and unlaminated,
  narrow, thin-film magnetic stripes},'' \emph{Journal of Applied Physics},
  vol.~69, pp. 5611--5615, 1991.

\bibitem{El-Ghazaly2015}
A.~El-Ghazaly, R.~M. White, and S.~X. Wang, ``{Increasing ferromagnetic
  resonance frequency using lamination and shape},'' \emph{Journal of Applied
  Physics}, vol. 117, no.~17, pp. 502--506, 2015.

\bibitem{Chen2000}
L.~H. Chen, T.~J. Klemmer, K.~A. Ellis, R.~B. van Dover, and S.~Jin,
  ``{Soft-magnetic properties of Fe – Co – B thin films for
  ultra-high-frequency applications},'' \emph{Journal of Applied Physics},
  vol.~87, no.~9, pp. 1--4, 2000.

\bibitem{Lebedev2012}
G.~A. Lebedev, B.~Viala, T.~Lafont, D.~I. Zakharov, O.~Cugat, and J.~Delamare,
  ``{Converse magnetoelectric effect dependence with CoFeB composition in
  ferromagnetic/piezoelectric composites},'' \emph{Journal of Applied Physics},
  vol. 111, no.~7, p. 07C725, 2012.

\bibitem{El-Ghazaly2013}
A.~El-Ghazaly, J.~M. Mullenix, R.~M. White, and S.~X. Wang, ``{Kerr-Imaged
  Edge-Curling Wall Effects of Narrow Magnetic Cores},'' \emph{IEEE
  Transactions on Magnetics}, vol.~49, no.~7, pp. 4017--4020, jul 2013.

\bibitem{Lopez-Villegas1998}
J.~Lopez-Villegas, J.~Samitier, and C.~Cane, ``{Improvement of the quality
  factor of RF integrated inductors bylayout optimization},'' \emph{IEEE
  Transacations on microwave theory and techniques}, vol.~48, no.~1, pp.
  76--83, 1998.

\bibitem{Meyer2010}
C.~D. Meyer, S.~S. Bedair, B.~C. Morgan, and D.~P. Arnold,
  ``{High-inductance-density, air-core, power inductors, and transformers
  designed for operation at 100-500 MHz},'' \emph{IEEE Transactions on
  Magnetics}, vol.~46, no.~6, pp. 2236--2239, 2010.

\bibitem{Raieszadeh2005}
M.~Raieszadeh, P.~Monajemi, S.-W. Yoon, J.~Laskar, and F.~Ayazi, ``{High-Q
  integrated inductors on trenched silicon islands},'' in \emph{IEEE
  International Conference on Micro Electro Mechanical Systems}, 2005, pp.
  199--202.

\bibitem{Frommberger2005}
M.~Frommberger, C.~Schmutz, M.~Tewes, J.~McCord, W.~Hartung, R.~Losehand, and
  E.~Quandt, ``{Integration of crossed anisotropy magnetic core into toroidal
  thin-film inductors},'' \emph{IEEE Transactions on Microwave Theory and
  Techniques}, vol.~53, no. 6 II, pp. 2096--2099, 2005.

\bibitem{Wang2016}
T.~Wang, Y.~Peng, W.~Jiang, Y.~M. Huang, B.~M.~F. Rahman, R.~Divan,
  D.~Rosenmann, and G.~Wang, ``{Integrating Nanopatterned Ferromagnetic and
  Ferroelectric Thin Films for Electrically Tunable RF Applications},''
  \emph{IEEE Transacations on microwave theory and techniques}, 2016.

\end{thebibliography}
\end{document}